\documentclass[11pt]{article}
\usepackage{moriond,epsfig}

\newcommand{\pt}{\rm{p}_{T}}
\newcommand{\Et}{\rm{E}_{T}}
\newcommand{\MET}{\mbox{$E\kern-0.50em\raise0.10ex\hbox{/}_{T}$}}

\newcommand {\GeV} {~\rm{GeV}\!/c^2}

\newcommand {\TeV} {~\rm{TeV}}
\newcommand {\pb} {~\rm{pb}^{-1}}
\newcommand {\fb} {~\rm{fb}^{-1}}

\begin{document}
\vspace*{4cm}
\title{DIBOSON PHYSICS AT THE TEVATRON}

\author{ M.S. NEUBAUER \\
(for the CDF and D\O\ Collaborations)}

\address{Department of Physics, University of California, San Diego, \\
La Jolla, California 92093}

\maketitle\abstracts{
At the Fermilab Tevatron, the CDF and D\O\ detectors are used to study
diboson production in $p\bar{p}$ collisions at $\sqrt{s}=1.96$ TeV. We
summarize recent measurements of the $W\gamma$, $Z\gamma$, and $WW$
cross sections and limits on $WZ$ and $ZZ$ production. Limits on
anomalous trilinear gauge couplings are also presented.
}

\section{Introduction}

The self-couplings of gauge bosons are a striking consequence of the
$\rm{SU(2)}_L\otimes\rm{U(1)}_Y$ structure of the Standard Model
(SM). Detailed study of diboson production provides a stringent test
of SM trilinear gauge couplings (TGC), which are sensitive to new
physics (NP) effects. These NP effects are parametrized as deviations
from SM couplings in an effective Lagrangian which, for TGC involving
two $W$ bosons, is given by \cite{Hagiwara:1993ck}
\begin{equation}
{\rm {\cal L}_{WWV} / g_{WWV}} = i {\rm g_1^V (W_{\mu\nu} W^\mu V^\nu -
W_\mu V_\nu W^{\mu\nu})} + i {\rm \kappa_V W_{\mu} W_\nu V^{\mu\nu}} +
\frac{i {\rm \lambda_{V}}}{{\rm M_W^2}} {\rm W_{\lambda\mu} W_\nu^\mu
V^{\nu\lambda}},
\end{equation}
where V$\equiv Z,\gamma$. This involves 5 C- and P-conserving coupling
parameters ($\rm{g_1^\gamma} = 1$ by EM gauge invariance) with SM values
at tree level given by $\Delta \kappa_Z=\Delta \kappa_\gamma=0
~(\Delta\kappa \equiv \kappa-1)$ and $\Delta g_1^Z = 0 ~ (\Delta g_1^Z \equiv
g_1^Z -1)$. Non-SM couplings increase the cross section at high
$\rm{E_T}$. A form factor ansatz is introduced to avoid unitarity
violation at large $\hat{s}$ such that $\alpha \rightarrow
\alpha(\hat{s}) = \alpha_0 / (1+\hat{s}/\Lambda_{\rm{FF}}^2)^2$ for a
given parameter $\alpha$. The NP causing non-SM couplings enter at a
scale $\Lambda_{\rm{FF}}$ assumed to be $\sim$1-2 TeV in the
analyses presented here.

In these proceedings, a brief summary of the current diboson results
from the CDF and D\O\ Collaborations at the Tevatron Run II are
presented. The diboson physics program at the Tevatron is
complementary to LEP since $p\bar{p}$ collisions at $\sqrt{s}=1.96\TeV$
probe different combinations of TGC couplings at higher $\hat{s}$,
where NP effects might become evident.

\section{$W\gamma$}

The $W\gamma$ final state observed at hadron colliders provides a
direct test of the $WW\gamma$ TGC. Anomalous $WW\gamma$ leads to an
enhancement in the production cross section and an excess of large $\Et$
photons. Both CDF and D\O\ have made measurements of the $W\gamma$ cross
section using leptonic decays of the $W$ bosons. The signature of
the $W\gamma$ signal is an isolated high $\Et$ lepton, an isolated
high $\Et$ photon, and large missing transverse energy ($\MET$) from
the $W$ neutrino. The dominant background is from $W$+jets where a jet
mimics an isolated photon. A lepton-photon separation requirement in
$\eta-\phi$ space of $\Delta R = \sqrt{(\Delta\eta)^2 +
(\Delta\phi)^2} > 0.7$ is made by both CDF and D\O\ to suppress events
with final-state radiation of the photon from the outgoing lepton and
to avoid collinear singularities in theory calculations. A kinematic
requirement on photon $\Et$ of $\Et > 7 (8) \GeV$ is made by CDF
(D\O\ ) in the analysis. 

CDF has published a measurement \cite{Acosta:2004it} with
$\int {\cal L}~dt = 200\pb$ of $\sigma(p\bar{p}\rightarrow
W\gamma + X) \times BR(W\rightarrow l\nu) = 18.1\pm1.6 \rm{(stat.)}
\pm 2.4 \rm{(syst.)} \pm 1.2 \rm{(lum.)}$ pb, to be compared with the
NLO theoretical expectation \cite{Baur:1992cd} of $19.3\pm1.4$ pb. D\O\
published a measurement \cite{Abazov:2005ni} with $\int {\cal L}~dt =
162\pb$ of $\sigma(p\bar{p}\rightarrow W\gamma + X) \times
BR(W\rightarrow l\nu) = 14.8\pm1.6 \rm{(stat.)} \pm 1.0 \rm{(syst.)}
\pm 1.0 \rm{(lum.)}$ pb to be compared with the NLO expectation
\cite{Baur:1992cd} of $16.0\pm0.4$ pb.

Both results are consistent with the SM expectations at NLO. D\O\ sets
limits on anomalous $WW\gamma$ TGC based upon the observed photon
$\Et$ spectrum. The one-dimensional limits at 95\% C.L. are
$-0.88<\Delta \kappa_\gamma<0.96$ and $-0.20<\lambda_\gamma<0.20$ for
$\Lambda_{\rm{FF}} = 2\TeV$. 

\section{$Z\gamma$}

In the SM, photons do not directly couple to $Z$ bosons at lowest
order. Therefore, observation of such a coupling would constitute
evidence for NP. The $Z\gamma$ final state at hadron colliders
involves a combination of $ZZ\gamma$ and $Z\gamma\gamma$
couplings. Both CDF and D\O\ have made measurements of the $Z\gamma$
cross section in leptonic decay channels of the $Z$ boson. The signature
of the $Z\gamma$ signal is two isolated high $\Et$ leptons having the
same flavor and opposite charge with invariant mass consistent with
decay of a $Z$ boson and an isolated high $\Et$ photon. The dominant
background is from $Z$+jets where a jet mimics an isolated photon. As in
the $W\gamma$ analyses, a lepton-photon separation requirement of
$\Delta R > 0.7$ is made by both CDF and D\O\. A kinematic requirement
on photon $\Et$ of $\Et > 7 (8) \GeV$ is made by CDF (D\O\ ) in the analysis.

CDF has published a measurement \cite{Acosta:2004it} with $\int {\cal
L}~dt = 200\pb$ of $\sigma(p\bar{p}\rightarrow
Z\gamma + X) \times BR(Z\rightarrow ll) = 4.6\pm0.5 \rm{(stat.)}
\pm 0.2 \rm{(syst.)} \pm 0.3 \rm{(lum.)}$ pb, to be compared with the
NLO theoretical expectation \cite{Baur:1992cd} of $4.5\pm0.3$ pb. D\O\
published a measurement \cite{Abazov:2005ea} with $\int {\cal L}~dt =
300\pb$ of $\sigma(p\bar{p}\rightarrow Z\gamma + X) \times
BR(Z\rightarrow ll) = 4.2\pm0.4 \rm{(stat.+syst.)} \pm 0.3 \rm{(lum.)}$
pb, to be compared with the NLO expectation \cite{Baur:1992cd} of
$3.9\pm0.2$ pb.

Both results are consistent with the SM expectations. D\O\ sets
anomalous coupling limits based upon the observed photon $\Et$
spectrum. The one dimensional limits at 95\% C.I. are
$|h_{10,30}^\gamma|<0.23$, $|h_{20,40}^\gamma|<0.019$,
$|h_{10,30}^Z|<0.23$, and $|h_{20,40}^Z|<0.020$ for $\Lambda_{\rm{FF}} =
1\TeV$. The limits on $|h_{20,40}^\gamma|$ and $|h_{20,40}^Z|$ are
the most stringent limits currently available. 

\section{$WW$}

Production of $W$ boson pairs involves both $WW\gamma$ and $WWZ$
couplings. At LEP, $WW$ production has been extensively studied and
stringent limits on anomalous TGC were determined. However, at the
Tevatron much higher $WW$ invariant masses are probed compared to LEP
because of the higher accessible energies. Also, the $WW$ final state
is a promising discovery channel for the Higgs boson at both the
Tevatron and the LHC. The signature of the $WW$ signal in
leptonic decay is two isolated high $\Et$ leptons with opposite charge
and large missing transverse energy ($\MET$) from the W
neutrinos. Events are required to have minimal jet activity to reduce
contamination from $t\bar{t}$ events. After the selection cuts, the
dominant backgrounds are from Drell-Yan, other diboson decays, and
$W$+jets where the jet fakes an isolated lepton. 

\begin{figure}
\begin{minipage}{0.50 \linewidth}
\begin{center}
\epsfig{file = 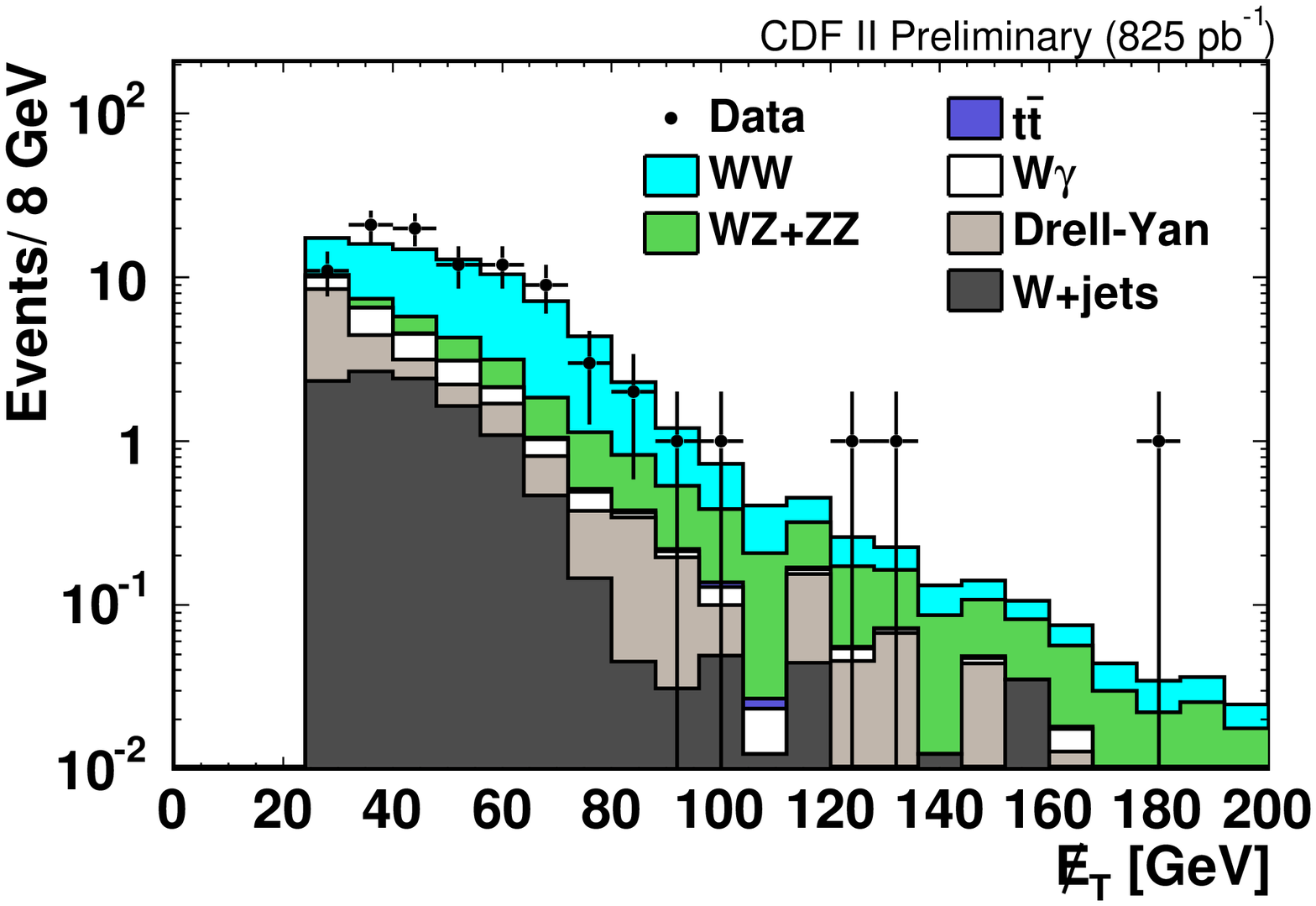,
width=0.99\linewidth}
\nolinebreak
\epsfig{file = 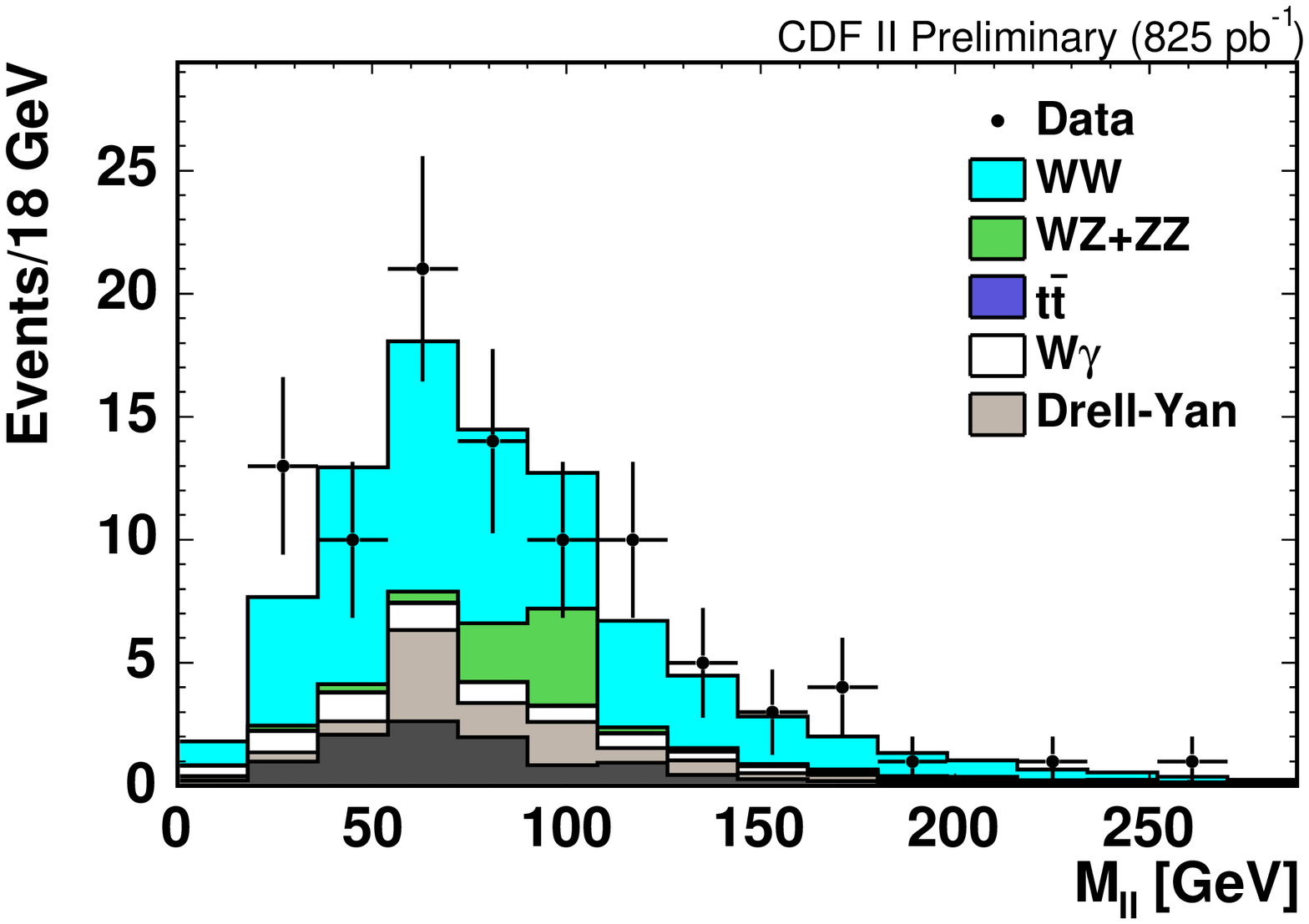, width = 0.99\linewidth}
\end{center}
\end{minipage}
\caption{$\MET$ (left) and dilepton mass (right) for $WW$ candidates.}
\label{fig:WW}
\end{figure}

D\O\ published a measurement \cite{Abazov:2004kc} in the dilepton channel
with $\int {\cal L}~dt = 240\pb$ of $\sigma(p\bar{p}\rightarrow WW +
X) = 13.8^{+4.3}_{-3.8} ~\rm{(stat.)} ^{+1.2}_{-0.9} ~\rm{(syst.)} \rm
0.9 ~\rm{(lum.)}$ pb. CDF has a new preliminary measurement with $\int
{\cal L}~dt = 825\pb$ of $\sigma(p\bar{p}\rightarrow WW + X) = 13.6\pm2.3
\rm{(stat.)} \pm 1.6 \rm{(syst.)} \pm 1.2 \rm{(lum.)}$ pb. In this
analysis, 95 events are observed with an expected signal (background)
of $52.4\pm4.4$ ($37.8\pm4.8$) events. Some kinematic distributions of
WW candidate events are shown in Figure~\ref{fig:WW}. The CDF
measurement is the most precise measurement of the $WW$ cross section
available from the Tevatron. Both measurements are consistent with the
NLO expectation \cite{Campbell:1999ah} of $12.4\pm0.8$ pb.

\section{$WZ/ZZ$}

Production of $WZ$ involves the $WWZ$ TGC. The production is
unavailable at LEP and has not been conclusively observed. The study
of $WZ$ production allows one to search for anomalous $WWZ$ coupling
independent of the $WW\gamma$ coupling, in contrast to $WW$
production. The NLO cross section for $WZ$ production at $\sqrt{s} =
1.96$ is $3.7\pm0.1$ pb \cite{Campbell:1999ah}. CDF
published \cite{Acosta:2005pq} a search for the sum of $WZ$ and $ZZ$
production in 2, 3, and 4 lepton channels with $\int {\cal L}~dt =
194\pb$ and set a limit of $\sigma(p\bar{p}\rightarrow WZ+ZZ) < 15.2$
pb at 95\% C.L. The SM expectation for $\sigma(p\bar{p}\rightarrow
WZ+ZZ)$ is $5.0\pm0.4$ pb.

\begin{figure}
\begin{minipage}{1.0 \linewidth}
\begin{center}
\epsfig{file = 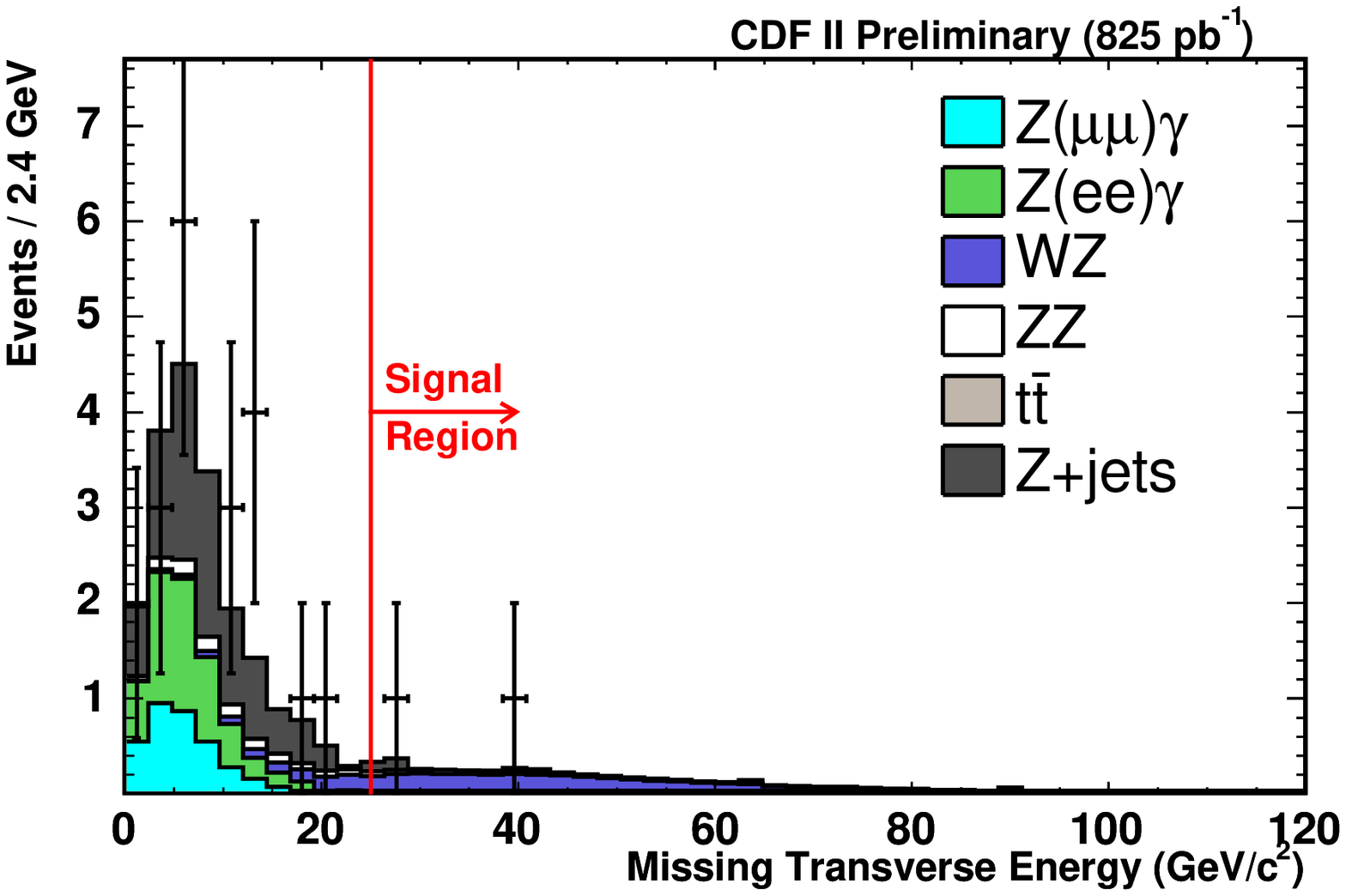, width = 0.99\linewidth}
\epsfig{file = 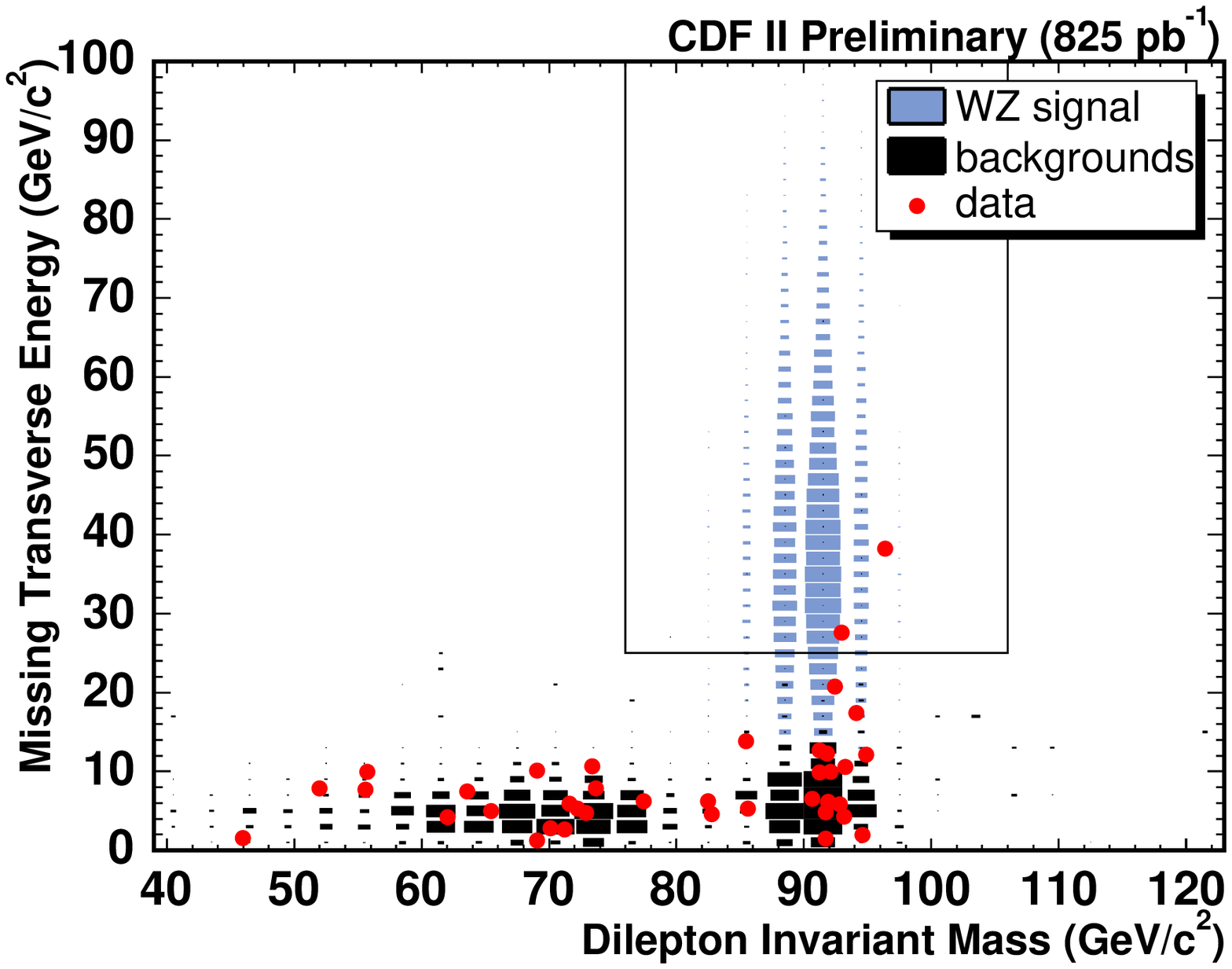, width = 0.99\linewidth}
\end{center}
\end{minipage}
\caption{(top) $\MET$ and (bottom) $\MET$ vs. dilepton mass for
trilepton candidates.}
\label{fig:WZ}
\end{figure}

CDF and D\O\ have made direct searches for $WZ$ production in the
trilepton + $\MET$ channel, assuming SM $ZZ$ production. The dominant
backgrounds are from $Z$+X, where X is a $Z$, $\gamma$, or jet faking
a lepton. The published D\O\ analysis \cite{Abazov:2005ys} used
$\int {\cal L}~dt = \sim300\pb$ and observed three events (1 eee,
2$\mu\mu\mu$) with an expected signal of $2.0\pm0.2$ events and
background of $0.71\pm0.08$ events. Based upon these results, they
quoted both an upper limit of $\sigma(p\bar{p}\rightarrow WZ) < 13.3$
pb at 95\% C.L. and a measurement of the cross section of
$\sigma(p\bar{p}\rightarrow WZ) = 4.5^{+3.8}_{-2.6} {\rm
(stat.+syst.)}$ pb. They also set limits on $WWZ$ anomalous TGC:
one-dimensional 95\% C.L. are $-2.0<\Delta \kappa_Z<2.4$ for
$\Lambda_{\rm{FF}} = 1.0\TeV$ and $-0.48<\lambda_Z<0.48$,
$-0.49<\Delta g_1^Z<0.66$ for $\Lambda_{\rm{FF}} = 1.5\TeV$. CDF has a
new preliminary result in the same decay channel with $\int {\cal
L}~dt = 825\pb$ where they observe two events (both eee) with an
expected signal of $3.7\pm0.3$ events and background of $0.9\pm0.2$
events. Based upon these results, they quote an upper limit of
$\sigma(p\bar{p}\rightarrow WZ) < 6.34$ pb at 95\%
C.I. Figure~\ref{fig:WZ} shows the $\MET$ and dilepton mass for
candidates both inside and outside the $WZ$ signal region.
\twocolumn

\section{$WW+WZ$ in \cal{l}$\nu$jj}

CDF has a new preliminary search for the sum of $WW$ and $WZ$ production
in the decay channel $l\nu$jj with $\int {\cal L}~dt = 350\pb$. The
advantage of this mode over the purely leptonic channels is the larger
branching fraction to jets, at the expense of larger backgrounds,
mainly from $W$+jets. Fitting the expected signal and background dijet
mass shape to data, the result shows no statistically significant
evidence for $WW+WZ$ production. A 95\% C.L. limit on the $WW+WZ$
cross section of 36 pb is determined from these results. Using the
observed $W~\pt$ spectrum, which is sensitive to anomalous TGC, 95\%
C.I. limits of $-0.51<\Delta \kappa<0.44$ and $-0.28<\lambda<0.28$ for
$\Lambda_{\rm{FF}} = 1.5\TeV$ are obtained.  

\section{Summary}

\begin{figure}
\begin{center}
\epsfig{file = 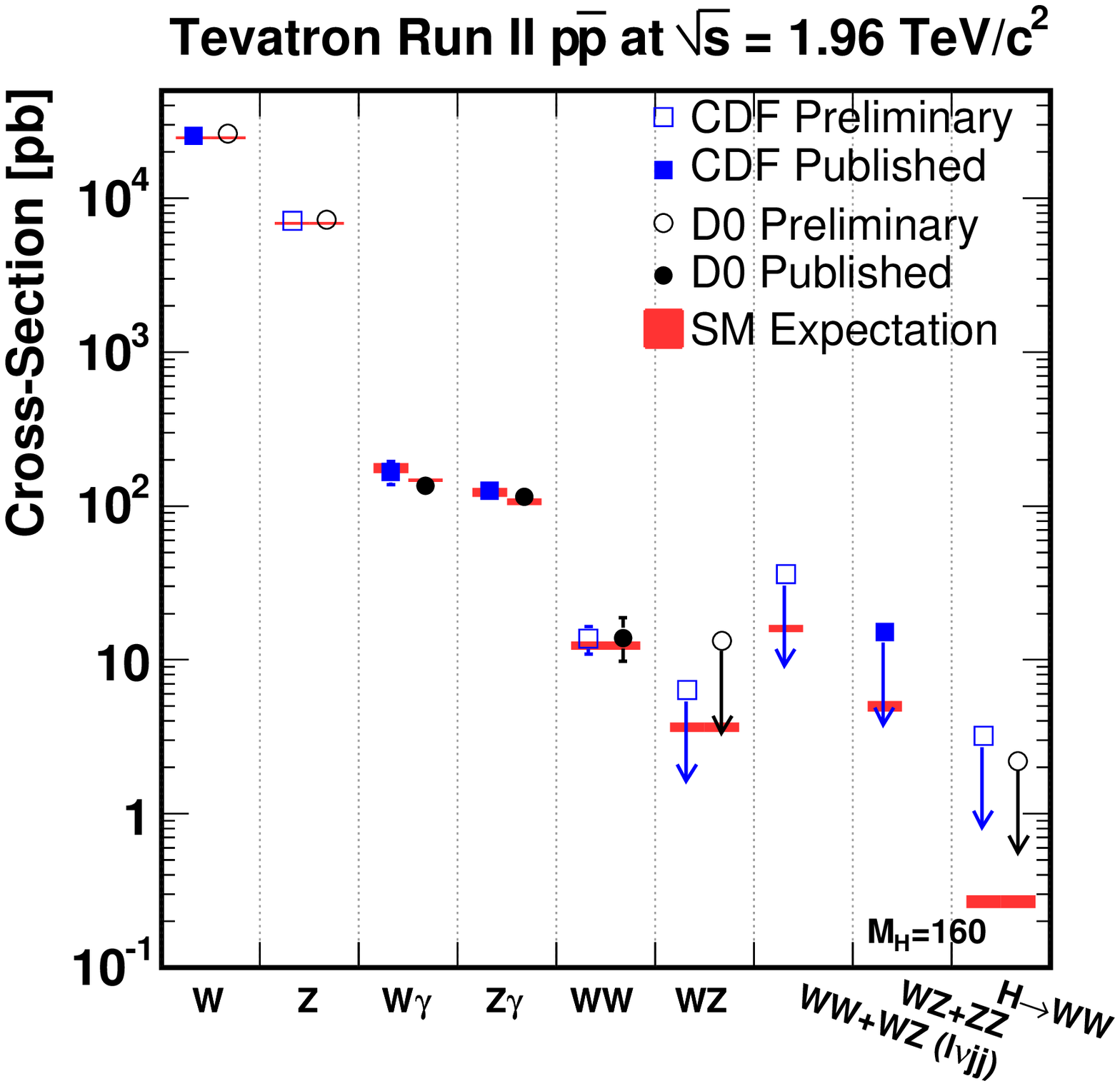, width=1.05\linewidth}
\end{center}
\caption{Electroweak cross section measurements from CDF and D\O\ .}
\label{fig:summary}
\end{figure}

Figure~\ref{fig:summary} summarizes the single boson and diboson cross
section measurements from the Tevatron. No deviation from the SM is
observed. With 4-8$\fb$ expected by the end of Run II, the electroweak
sector of the SM will be put to more stringent tests in the search for
NP.

\bibliography{moriond}

\end{document}